\documentclass{llncs}

\usepackage[T1]{fontenc}
\usepackage[utf8]{inputenc}

\usepackage{graphicx}
\usepackage[caption=false]{subfig}

\newcommand{\tre}[1]{{\tt #1}}

\usepackage[cmex10]{amsmath}

\usepackage{textcomp}
\usepackage{graphicx}
\usepackage{color}
\usepackage{tikz}
\usetikzlibrary{arrows,shapes,decorations.text,decorations.pathreplacing}

\definecolor{material indigo}{HTML}{3F51B5}
\definecolor{material teal}{HTML}{009688}
\definecolor{material blue grey}{HTML}{607D8B}
\definecolor{material green}{HTML}{4CAF50}
\definecolor{material red}{HTML}{FF5252}
\definecolor{material green}{HTML}{4CAF50}
\definecolor{material yellow}{HTML}{FFEB3B}
\definecolor{material amber}{HTML}{FFC107}
\definecolor{material lime}{HTML}{CDDC39}
\definecolor{material light green}{HTML}{8BC34A}
\definecolor{material deep orange}{HTML}{FF5722}
\definecolor{material orange}{HTML}{FF9800}
\definecolor{material blue}{HTML}{2196F3}
\definecolor{material pink}{HTML}{FF4081}
\definecolor{material purple}{HTML}{9C27B0}
\definecolor{material deep purple}{HTML}{7C4DFF}
\definecolor{material light blue}{HTML}{03A9F4}
\definecolor{material cyan}{HTML}{00BCD4}
\definecolor{material brown}{HTML}{795548}
\definecolor{material grey}{HTML}{9E9E9E}

\usepackage{booktabs}

\usepackage[firstpage]{draftwatermark}\SetWatermarkText{\hspace*{5.2in}\raisebox{5in}{\includegraphics[scale=0.1]{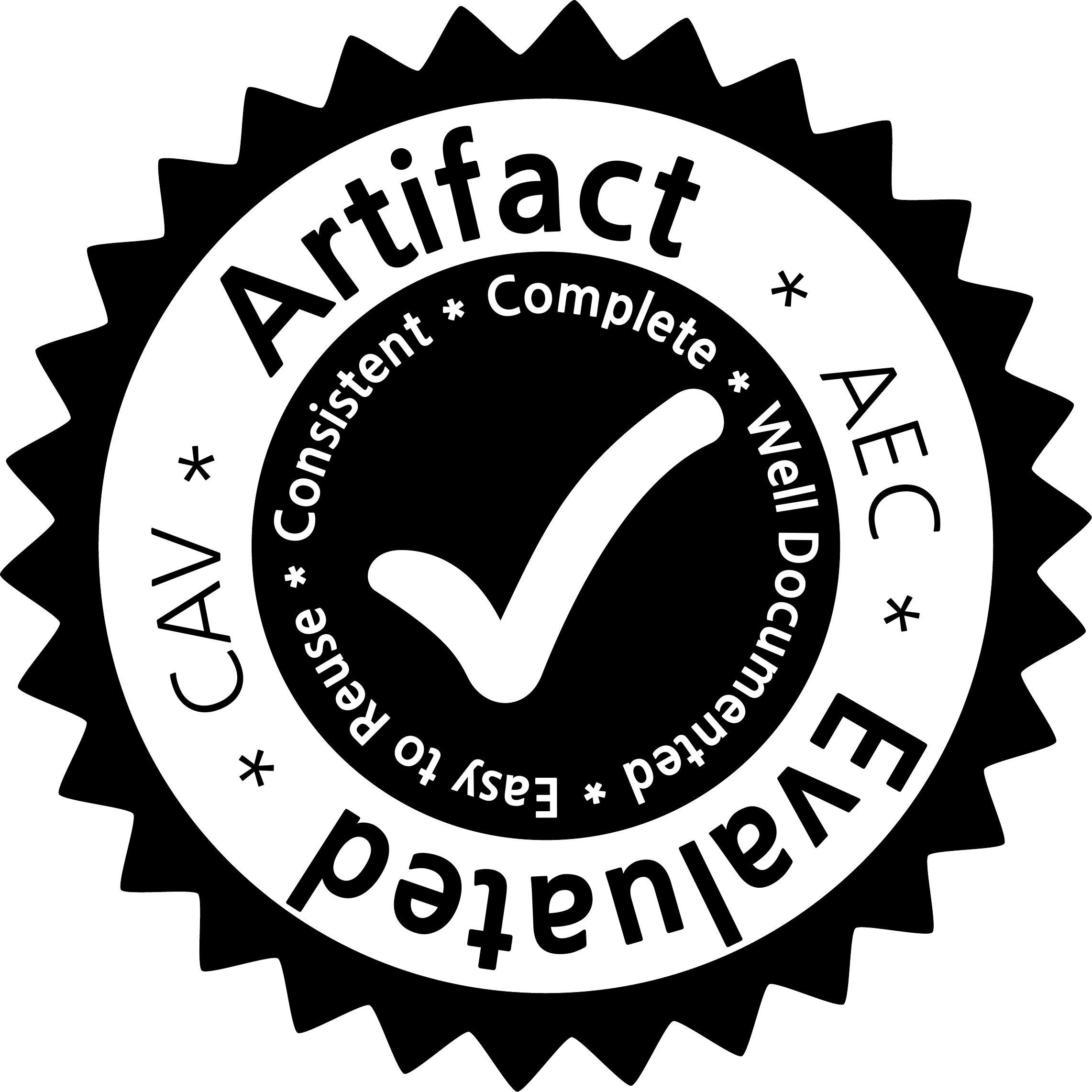}}}\SetWatermarkAngle{0}

\everymath{\displaystyle}

\title{{\sc Montre}: A Tool for\\ Monitoring Timed Regular Expressions}

\author{Dogan Ulus}
\institute{{\sc Verimag}, Universit\'e Grenoble-Alpes, Grenoble, France}

\begin{document}

\maketitle

\begin{abstract}
We present {\sc Montre}, a monitoring tool to search patterns specified by timed regular expressions over real-time behaviors.
We use timed regular expressions as a compact, natural, and highly-expressive pattern specification language for monitoring applications involving quantitative timing constraints.
Our tool essentially incorporates online and offline timed pattern matching algorithms so it is capable of finding all occurrences of a given pattern over both logged and streaming behaviors.
Furthermore, {\sc Montre} is designed to work with other tools via standard interfaces to perform more complex and versatile tasks for analyzing and reasoning about cyber-physical systems.
As the first of its kind, we believe {\sc Montre} will enable a new line of inquiries and techniques in these fields.
\end{abstract} %

\section{Introduction}

Temporal behaviors are sequences of actions and observations in time generated by various systems and the environment around us.
A temporal pattern is a set of compositions of different temporal behaviors satisfying some relations among their components such as precedence and coincidence or possessing some properties such as repetition and a certain duration.
Searching good [bad, desirable, undesirable] patterns over their temporal behaviors is an important task while we reason about systems and the environment.

Timed regular expressions (TREs)~\cite{tre} extend regular expressions, a well-established formalism for specifying sequences of symbols, with the notion of real-time and timing constraints.
Many patterns requiring both qualitative and quantitative temporal properties can be specified by TREs in a compact and natural way.
Given a TRE that specifies a temporal pattern and a real-time behavior the problem of timed pattern matching is defined as locating all segments that satisfy the expression. This problem has been solved by an offline algorithm  in~\cite{patterns}.
It is further endowed with an online algorithm that incrementally matches patterns over streaming behaviors~\cite{timed-deriv}.

In this paper, we describe {\sc Montre} a new tool for timed pattern matching whose applications are numerous and diverse.
First of all, {\sc Montre} can naturally check execution traces of software and hardware systems against real time properties specified in TRE (e.g. ~\cite{measlang,pattern-elastic}), thus complementing temporal logic based property checkers such as~\cite{amt,breach,staliro}.
Further, {\sc Montre} can be used for specification mining such as~\cite{mining-stl,mining-tre-sebastian} as matching is a basic task for mining.
Outside the verification context, {\sc Montre} has a potential use in temporal data mining~\cite{mitsa2010temporal} and (vehicle or human) trajectory data mining~\cite{zheng2015trajectory,mazimpaka2016trajectory} as it can label time segments with meaningful tags such as overtaking (another car) or sprinting.
To illustrate our tool in action, we present such an example from the domain of sports analytics in Section~\ref{sec:ex} where we find all sprints of a soccer player.

\section{Tool Description}
\label{sec:io}

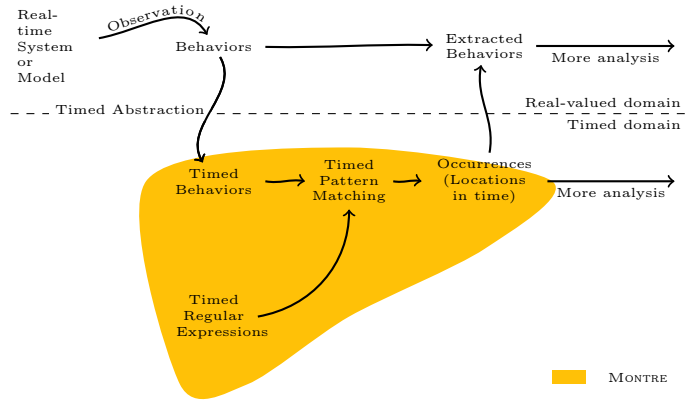
\begin{figure}[tb]
\centering
\begin{tikzpicture}[scale=0.9]
\tiny
\draw [dashed] (0,0) -- (10,0);
\node [rectangle, anchor=south east] (deep) at (10,0) {\parbox[c]{3cm}{\raggedleft\tiny Real-valued domain}};
\node [rectangle, anchor=north east] (time) at (10,0) {\parbox[c]{2cm}{\raggedleft\tiny Timed domain}};
\node [rectangle, anchor=west] (sys) at (0,1) {\parbox[c]{1cm}{Real-time\\System\\or\\Model}};
\node [rectangle] (beh) at (3,1) {\parbox[c]{1.2cm}{\centering Behaviors}};
\draw [thick, ->, postaction={decorate,decoration={text along path,text align=center, raise=0.5mm,text={Observation}}}] (sys) to[out=10,in=120] (beh);

\fill[material amber] plot [smooth cycle] coordinates {(2,-1) (2.5,-4) (3.5,-4) (5, -3) (7, -2) (8,-1) (5,-0.5)};
\node [rectangle] (sig) at (3,-1) {\parbox[c]{1.2cm}{\centering Timed Behaviors}};
\draw [thick, ->] (beh) to[out=-60,in=120] (sig);
\draw [->] (beh) to[out=-60,in=120] node[left, fill=white] {Timed Abstraction\quad} (sig);

\draw [thick, ->] (beh) to[out=-60,in=120] (sig);
\node [rectangle] (tre) at (3,-3) {\parbox[c]{1cm}{\centering Timed\\Regular\\Expressions}};
\node [rectangle] (tpm) at (5,-1) {\parbox[c]{1cm}{\centering Timed\\Pattern\\Matching}};
\node [rectangle] (match) at (7,-1) {\parbox[c]{1.5cm}{\centering Occurrences\\(Locations\\ in time)}};
\draw [thick, ->] (sig.east) to[out=-10,in=170] (tpm.west);
\draw [thick, ->] (tre.east) to[out=10,in=-90] (tpm.south);
\draw [thick, ->] (tpm.east) to[out=-10,in=170] (match.west);

\fill[material amber] (8,-3.8) rectangle (8.5,-4);
\node [rectangle, anchor=west] (legend) at (8.5,-3.9) {\parbox[c]{1.2cm}{\centering\tiny\sc Montre}};

\node [rectangle] (ibeh) at (7,1) {\parbox[c]{1.2cm}{\centering Extracted\\Behaviors}};
\node [rectangle, anchor=east] (tam) at (10,-1) {};
\node [rectangle, anchor=east] (ram) at (10,1) {};
\draw [thick, ->] (match) to[out=80,in=-100] (ibeh);
\draw [thick, ->] (match) -- node[below] {\tiny More analysis} (tam);
\draw [thick, ->] (ibeh) -- node[below] {\tiny More analysis} (ram);
\draw [thick, ->] (beh) to[out=-1,in=179] (ibeh);

\end{tikzpicture}
\caption{The work flow and extent of the monitoring tool {\sc Montre}}
\label{fig:montre}
\end{figure}

The tool {\sc Montre} essentially incorporates online and offline timed pattern matching algorithms extended with some practical features such as anchors and a Boolean layer.
It takes a timed behavior and a timed regular expression as inputs, and produces a finite set of two dimensional zones representing the (possibly uncountable) set of segments that watch the pattern.
{\sc Montre} provides a standard text-based interface for easy integration with other tasks such as data preparations and visualization as we consider them necessary but outside the scope of {\sc Montre}.
In Figure~\ref{fig:montre}, we illustrate the work flow and extent of {\sc Montre}, and we give details for each component in the following.

\subsubsection{Timed behaviors.}
A timed behavior is a sequence of time segments where each segment has a duration value and is associated with a set of propositional variables that hold continuously in the segment.
In general, we assume all propositions are concurrent.
For example, {\tt (3,pq);(2,q);(2,p)} is a timed behavior with 3 segments over propositions {\tt p} and {\tt q}.
It means that {\tt p} and {\tt q} evaluate to true for the first 3 time units, then {\tt q} is true for 2 more time units, and then {\tt p} is true for 2 time units again.
We assume behaviors start at time $0$; therefore, the example behavior can be alternatively stated such that {\tt p} holds from 0 to 3 and then 5 to 7 while {\tt q} holds from 0 to 5.

\subsubsection{Timed regular expressions.}
An atomic timed regular expression corresponds to a Boolean expression over a set of propositions, denoted by letters {\tt p}, {\tt q}, {\tt r}.
These propositions can stand for predicates over real-valued variables.
Usual Boolean operators ({\tt !}), ({\tt ||}), ({\tt \&\&}) are used to build Boolean expressions.
We say that an atomic expression occurs on a time period $(t,t')$ if the corresponding Boolean expression holds from $t$ to $t'$ continuously.
Complex timed regular expressions are built from other expressions by using {\sc tre} operators: sequential composition (concatenation) (\tre{;}), time restriction (\tre{\%}), choice (\tre{|}), coincidence (\tre{\&}) and zero-or-more repetition (\tre{*}).
Further, we add one-or-more repetition (\tre{+}) and two anchoring (\tre{<:} and \tre{:>}) operators to the set of operators.
Typically parentheses are used to group expressions.
We summarize all Boolean and {\sc tre} operations in {\sc Montre} in Table~\ref{tbl:syntax}.

\begin{table}[t]
\caption{{\sc Montre} Timed Regular Expression Syntax}
\centering
\begin{tabular}{cp{9.5cm}r}

\toprule
Construct & \centering Description & \\
\midrule
{\tt p} & A propositional variable.   & \\
{\tt !P} & Boolean {\sc not} operation on {\tt P}.   & \\
{\tt P || Q} & Boolean {\sc or} operation on {\tt P} and {\tt Q}. & \\
{\tt P \&\& Q} & Boolean {\sc and} operator on {\tt P} and {\tt Q}.  & \\
\midrule
{\tt P} & occurs on $(t, t')$ if {\tt P} holds from $t$ to $t'$ continuously. & \\
{\tt <:P} & occurs on a time period $(t, t')$ if {\tt P} occurs on  $(t, t')$ and there exists a rising edge for {\tt P} at $t$. & \\
{\tt P:>} & occurs on a time period $(t, t')$ if {\tt P} occurs on  $(t, t')$ and there exists a falling edge for {\tt P} at $t'$.   & \\
{\tt <:P:>} & occurs on a time period $(t, t')$ if {\tt P} occurs on  $(t, t')$ and there exists a rising edge for {\tt P} at $t$ as well as a falling edge for $P$ at $t'$. & \\
{\tt E;F} & occurs on a time period $(t, t')$ if {\tt E} occurs on $(t, t'')$ and {\tt F} occurs on $(t'', t')$ for $t \leq t'' \leq t'$. & \\
{\tt E|F} & occurs on a time period $(t, t')$ if either {\tt E} or {\tt F} occurs on $(t, t')$. & \\
{\tt E\&F} & occurs on a time period $(t, t')$ if {\tt E} and {\tt F} occur on $(t, t')$ concurrently. & \\
{\tt E\%(m,n)} & occurs on a time period $(t, t')$ if {\tt E} occurs on $(t, t')$ and the duration of the occurrence is in the specified range such that {\tt m} $\leq t'-t \leq$ {\tt n}. & \\
{\tt E*} & Zero-or-more repetition of {\tt E}. & \\
{\tt E+} & One-or-more repetition of {\tt E}. & \\

\bottomrule

\end{tabular}
\label{tbl:syntax}
\end{table}
\vspace{-0.5cm}
\subsubsection{Zones.}
\begin{figure}[t]
\centering
\begin{tikzpicture}[scale=0.4]
\draw[very thin, material grey] (0, 0) grid (6, 6);
\fill[white] (0,0) -- (6.1,6.1) -- (6.1,0) --cycle;

\draw[->] (0,0) -- (0,7) node[anchor=east] {\tiny End};
\draw[->] (0,0) -- (7,0) node[anchor=north west] {\tiny Begin};
\fill[material amber, fill opacity=.9] (1,1) -- (1,3) -- (3, 3) -- cycle;
\draw[->] (0,0) -- (6,6);
\draw[material red, thick] (1,0) -- (1,6);
\draw[material red, thick] (0,3) -- (6,3);

\node[anchor=north west] (b) at (1,3) {\tiny $Z_{1}$};
\node[anchor=north] (b) at (1,0) {\tiny $t_{1}$};
\node[anchor=east] (b) at (0,3) {\tiny $t_{2}$};

\node (i) at (3.65, -3.4) {(i)};

\phantom{\node (i) at (3.65, -1.7) {\parbox[c]{1.8cm}{\centering\tiny
$b \leq t_{1} < b'$\\
$e < t_{2} \leq e'$\\
$d \leq t_{2}-t_{1} \leq d'$
}};}
\node (i) at (3.65, -1.7){\tiny $Z_1: t_{1}\leq t < t' \leq t_{2}$};

\end{tikzpicture}
\begin{tikzpicture}[scale=0.4]
\draw[very thin, material grey] (0, 0) grid (6, 6);
\fill[white] (0,0) -- (6.1,6.1) -- (6.1,0) --cycle;
\draw (0,0) -- (6,6);
\draw[->] (0,0) -- (0,7) node[anchor=east] {\tiny End};
\draw[->] (0,0) -- (7,0) node[anchor=north west] {\tiny Begin};
\fill[material amber, fill opacity=.9] (1,3) -- (1, 4.5) -- (1.5, 5) -- (3,5) -- (3, 4) -- (2,3) -- cycle;

\draw (0,0) -- (6,6);
\node[anchor=north] (b) at (1,0) {\tiny$b$};
\node[anchor=north] (b) at (3,0) {\tiny$b'$};
\node[anchor=east] (b) at (0,3) {\tiny$e$};
\node[anchor=east] (b) at (0,5) {\tiny $e'$};
\node[anchor=east] (b) at (0,3.5) {\tiny$d'$};
\node[anchor=east] (b) at (0,1) {\tiny$d$};
\node[] (b) at (2,4) {\tiny $Z_{2}$};

\node (i) at (3.65, -1.7) {\parbox[c]{1.8cm}{\centering\tiny
$b \leq t < b'$\\
$e < t' \leq e'$\\
$d \leq t'-t \leq d'$
}};

\node (i) at (1.4, -1.7) {\tiny $Z_2:$};

\draw[material red, thick] (1,0) -- (1,6);
\draw[material red, thick, dashed] (3,0) -- (3,6);
\draw[material red, thick, dashed] (0,3) -- (6,3);
\draw[material red, thick] (0,5) -- (6,5);
\draw[material red, thick] (0,3.5) -- (2.5,6);
\draw[material red, thick] (0,1) -- (5,6);

\node (i) at (3.65, -3.4) {(ii)};

\end{tikzpicture}
\caption{(i) A triangular zone.\quad (ii) A zone in general.}
\label{fig:zone}
\end{figure}
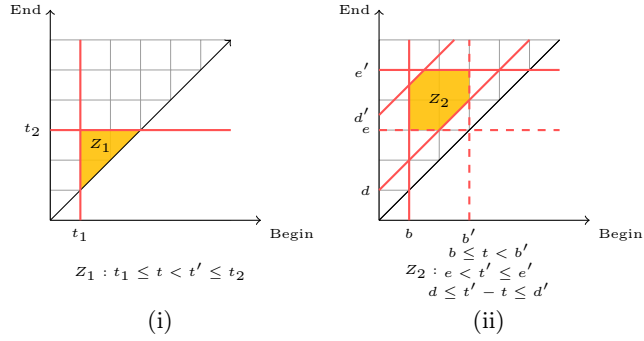

For a proposition {\tt p} that holds from $t_{1}$ to $t_{2}$, all sub-periods of $(t_1, t_2)$ satisfy the expression {\tt p}.
As shown in Figure~\ref{fig:zone}-(i), such a set of matches $\{(t, t')\ |\ t_{1}\leq t < t' \leq t_{2}\}$ can be represented on a two-dimensional plane as a triangular zone.
Then the match set of any atomic expression would be a union of such triangular zones.
A triangular zone is a special case of zones, which constitutes a restricted class of convex polygons defined by orthogonal and diagonal constraints as shown Figure~\ref{fig:zone}-(ii).
Zones are basic data objects for timed pattern matching as unions of zones are closed under Boolean and regular operations.
It follows that the match set of any timed regular expression over a timed behavior can be representable by a finite union of zones.

\subsubsection{Implementation.}
\label{sec:tool}

{\sc Montre} is a command line program\footnote{Available at http://github.com/doganulus/montre} that uses structured text files for input/output specification.
When invoked {\sc Montre} parses the timed regular expression passed as an argument and starts to reads the input file.
According to flags set by the user {\sc Montre} would run in either online or offline mode.
For online mode it is possible that the input can be given interactively using the command line or directed from another process as usual.
At its core, {\sc Montre} contains our efficient zone manipulation library, {\tt libmontre}, called dynamically by top-level online and offline timed pattern matching algorithms.
As Boolean and regular operations over sets of zones are intensive numerical computations, we have implemented {\tt libmontre} in C++.
In the implementation, we use an integer-valued time model where all time values are represented by integers for efficiency and accuracy reasons.
For the majority of applications, integers give us sufficient precision and range; and a proper scaling can be found.

We implement timed pattern matching algorithms in Pure\footnote{Available at http://purelang.bitbucket.io}, a functional programming language based on term rewriting  with a support for native code compilation and native
calls to dynamic libraries.
For the online algorithm~\cite{timed-deriv}, built upon derivatives of regular expressions~\cite{rederiv,rewrite-ere}, we extensively use the rewriting functionality when deriving an expression with respect to a newly observed segment.
The offline algorithm~\cite{patterns} is a recursive computation over the syntax tree of the expression; therefore, the role of Pure's rewriting engine is minimal.
The worst case complexity is polynomial in the size of input behavior and expression for the offline approach.
For the online approach it is polynomial in the size of the behavior and exponential in expression.
In practice, however, we realistically assume patterns to be much shorter than behaviors and somewhat sparse in them.
Then we expect a linear-time performance in the size of input behavior for both algorithms.
Under these assumptions, {\sc Montre} can process timed behaviors with a size of 1M segments in a few seconds (offline) and a few hundred seconds (online).

\section{An Illustrative Example}
\label{sec:ex}

We present an example use of Montre on a data set obtained by tracking positions of players in a real soccer match.
In this example, we find all sprints performed by a single player where a sprint is formally specified by a timed regular expression over speed and acceleration behaviors.
%
%
The data are obtained by a computer vision algorithm with a frame rate of 10 Hz so we have a $xy$-coordinate for each player on the field at every 100 milliseconds.
Therefore we use milliseconds as our base time unit for behaviors and expressions.

\begin{table}[b]
\centering
\caption{Speed and acceleration thresholds~\cite{sprintdef}.}
\begin{tabular}{ccc}
\begin{tabular}{ccc}
\toprule
Symbol & Label & \parbox{3cm}{\centering Speed \\thresholds ($m\cdot s^{-1}$)}\\
\midrule
{\tt s} & High & > 6.0\\
{\tt r} & Medium & 3.7 - 6\\
{\tt q} & Low & 2 - 3.7\\
{\tt p} & Near Zero & 0 - 2\\
\bottomrule
\end{tabular}
&\quad\quad&
\begin{tabular}{ccc}
\toprule
Symbol & Label & \parbox{3cm}{\centering Acceleration \\thresholds ($m\cdot s^{-2}$)}\\
\midrule
{\tt g} & High & >1.60\\
{\tt f} & Medium & 1.17 - 1.60\\
{\tt e} & Low & 0.57 - 1.17\\
{\tt d} & Near Zero & -0.57 - 0.57\\
\bottomrule
\end{tabular}
\end{tabular}
\label{tbl:categories}
\end{table}

In order to specify a pattern for sprints, we need to address two issues in order: (1) how to categorize continuous speed and acceleration axes, and (2) which composition of these categories defines a sprinting effort best.
Clearly, there are no universal answers for these questions so we rely on the study~\cite{sprintdef} in the following.
First, we partition speed and acceleration axes into four categories (near-zero, low, medium, and high), and we associate a letter for each category in Table~\ref{tbl:categories}.
For example, a period of medium speed, denoted by {\tt r}, means the speed value resides between $3.7$ and $6$ m/s during the period.

Often a sprint effort is characterized by any movement above a~certain speed threshold for a limited time.
This gives us our first sprint pattern such that a~period of high speed between 1-10 seconds, formally written as follows:
\begin{equation}
	\text{\tt (<:s:>)\%(1000,10000)} \tag{{\tt P1}}
\end{equation}
Above we use anchor operators from both sides on the proposition {\tt s} to obtain only maximal periods that satisfy {\tt s}; otherwise, any sub-period satisfies the pattern as well.
The operator {\tt\%} specifies that the duration is restricted to be in 1000 and 10000 milliseconds.
Alternatively we may want to find other efforts starting with high acceleration but not reaching top speeds necessarily.
This gives us our second sprint pattern such that a~period of high acceleration followed by a~period of medium or high speed between 1-10 seconds, formally written as follows:
\begin{equation}
	\text{\tt (<:g);(<:(r||s):>)\%(1000,10000)} \tag{{\tt P2}}
\end{equation}
Notice that we do not use the right-anchor on {\tt g}.
This allows a medium or high speed period to overlap with a high acceleration period as it is usually the case that they are concurrent.
Writing an equivalent pattern using classical regular expressions over a product alphabet would be a very tedious task partly due to a requirement to handle such interleavings explicitly (and the lack of timing constraints).
For TREs all propositions are considered to be concurrent by definition, which results in concise and intuitive expressions.
\begin{figure}[t]
\begin{center}
\subfloat[Entire movement]{\includegraphics[width=0.36\textwidth, height=0.14\textheight]{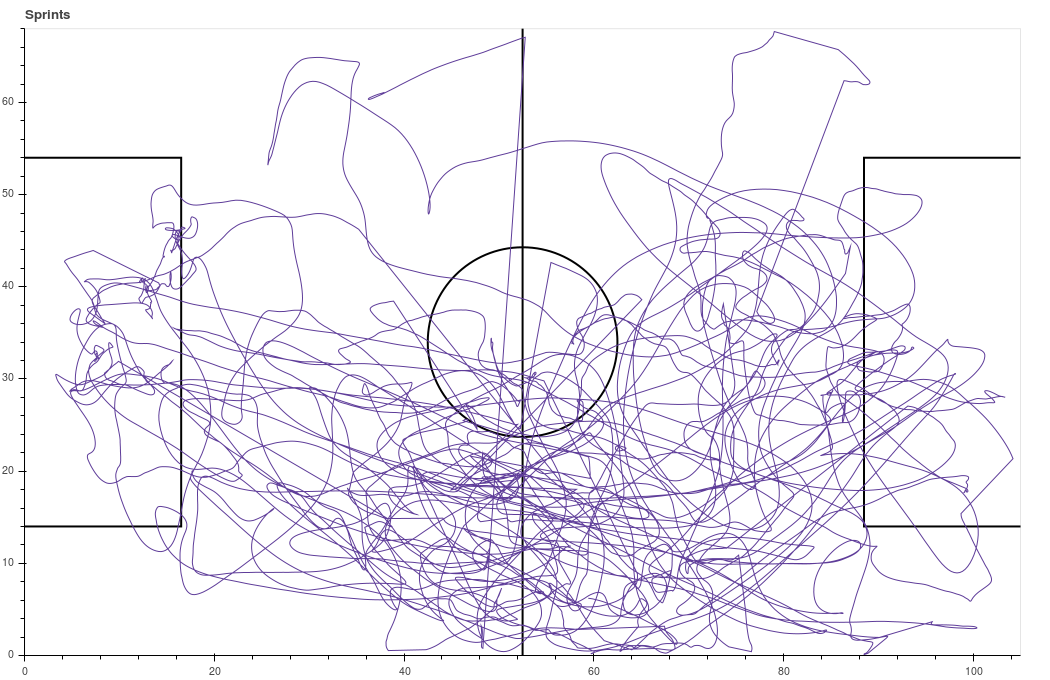}}
\subfloat[{\tt P1}]{\includegraphics[width=0.36\textwidth,height=0.14\textheight]{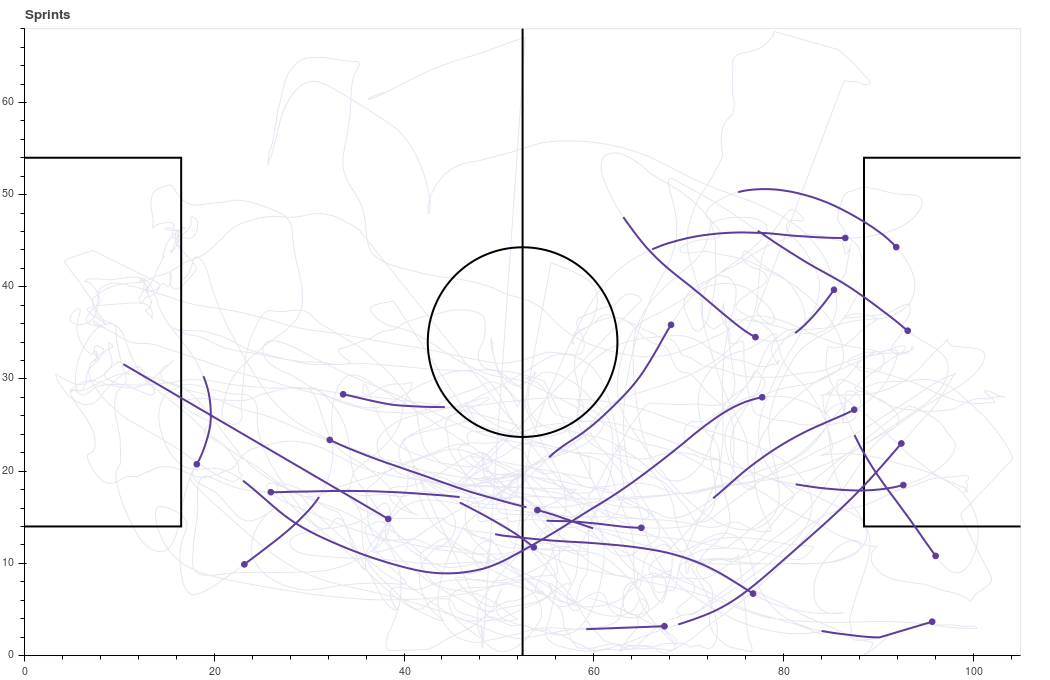}}
\\
\noindent
\subfloat[{\tt P2}]{\includegraphics[width=0.36\textwidth, height=0.14\textheight]{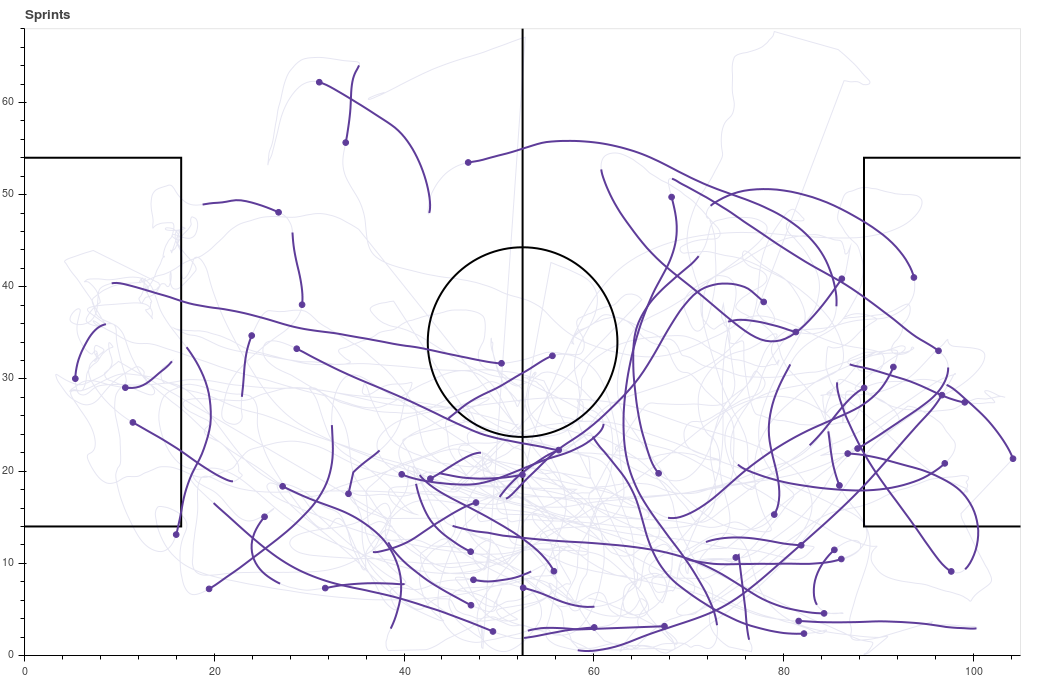}} 
\subfloat[{\tt P3}]{\includegraphics[width=0.36\textwidth,height=0.14\textheight]{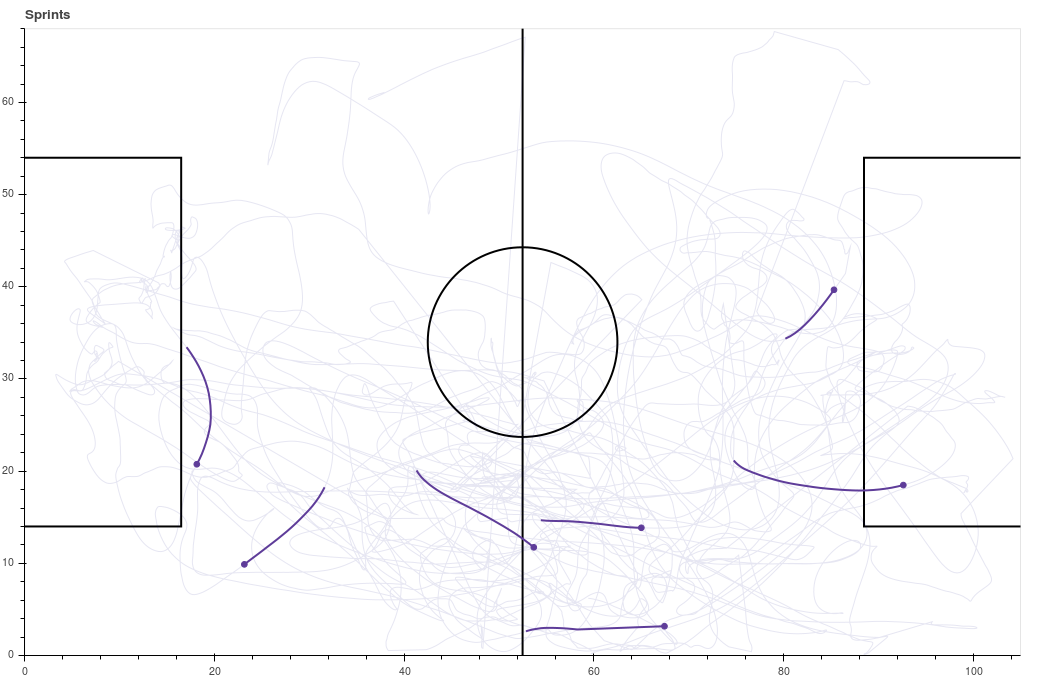}}
\end{center}
\caption{The trajectory of a soccer player for 45 minutes on the field, and his sprinting periods found by {\sc Montre} for patterns {\tt P1-P3}.}
\label{fig:sprints}
\vspace{-0.5cm}
\end{figure}
Finally we give a third pattern to find rather short but intense sprints such that
\begin{align*}
	\text{\tt (<:(f||g));((<:s:>)\%(1000,2000))} \tag{{\tt P3}}
\end{align*}
Then we visualize all sprints found by {\sc Montre} for patterns {\tt P1-P3} in Figure~\ref{fig:sprints} over the behavior of a single player during one half of the game (45 min.) containing 27K data points that reduces to timed behaviors of 5K segments after pre-processing.
Note that we used Python to prepare data and visualize results.

\section{Conclusions}

Timed regular expressions can define many timed properties and {\sc Montre} is the first tool to check such properties and detect timed patterns.
Its performance is satisfactory for such monitoring tasks but we note that there is still some room for optimization especially for the online algorithm.
The example we presented illustrates a complete {\sc Montre} experience from raw data to visualization.
As seen defining good patterns and categories are important to achieve intended results but it is not always obvious what a good pattern is.
Such patterns should be found in the future using (unsupervised) pattern mining methods.
We believe {\sc Montre} would provide a good starting point for such research as it encapsulates timed pattern matching with an easy-to-use interface.

\noindent\textbf{Acknowledgment.} Thanks to Oded Maler for his helpful comments on the text, and to Hande Alemdar and Serdar Alemdar for the soccer data they provided.

\bibliographystyle{plain}
\bibliography{master}

\end{document}